\documentclass[%
twocolumn,
 amsmath,amssymb,
 aps,
]{revtex4-2}

\usepackage{graphicx}
\usepackage{dcolumn}
\usepackage{bm}
\usepackage{xcolor}
\usepackage{algorithm} 
\usepackage{algpseudocode} 

\begin{document}


\title{Copula-based analysis of the autocorrelation function for simple temporal networks}

\author{Hang-Hyun Jo}
\email{h2jo@catholic.ac.kr}
\affiliation{Department of Physics, The Catholic University of Korea, Bucheon 14662, Republic of Korea}

\date{\today}

\begin{abstract}
To characterize temporal correlations in temporal networks, we define an autocorrelation function (ACF) for temporal networks in terms of the similarity between two snapshot networks separated by a certain time interval. By employing a copula-based method recently developed for a single time series, we analyze the ACF for the temporal network in which activity patterns of links are independent of each other but their activity levels are heterogeneous. By assuming that exponential distributed interevent times are weakly correlated with each other in each link, we obtain an analytical solution of the ACF. The validity of the analytical solution is tested against the numerical simulations to find that the numerical results are comparable to the analytical solution.
\end{abstract}

\maketitle

\section{Introduction}

For the past decades, complex systems have often been studied in the framework of network science: elements of the system and their pairwise interactions are denoted by nodes and links, respectively~\cite{Barabasi2016Network, Newman2018Networks, Menczer2020First}. One of recently emerging topics in the network science is temporal networks, in which a link connecting two nodes is considered to exist only when those nodes interact with each other~\cite{Holme2012Temporal, Holme2019Temporal, Masuda2016Guide}. Such a temporal interaction pattern can be described by a sequence of interaction events, namely, an event sequence. In many real-world datasets, the event sequences show a non-Poissonian or bursty nature~\cite{Karsai2018Bursty}, meaning that rapidly occurring events in short time periods are alternated with long inactive periods~\cite{Barabasi2005Origin}. Bursty patterns observed in link activities as well as in node activities have been known to play a key role in understanding the structure of temporal networks as well as dynamical processes taking place in them such as spreading and diffusion~\cite{Masuda2017Random, Karsai2018Bursty, Holme2019Temporal, Hiraoka2020Modeling}. Therefore it is crucial to properly characterize bursty time series. 

To characterize bursty time series, a number of quantities and methods have been proposed~\cite{Karsai2018Bursty, Jo2019Bursty}. The most basic quantity must be a time interval between two consecutive events, which is called an interevent time (IET). Bursty patterns have been typically described by the heavy-tailed IET distribution, which however ignores correlations between IETs. The correlations between two consecutive IETs can be measured in terms of memory coefficient~\cite{Goh2008Burstiness}, while those between an arbitrary number of consecutive IETs at some given timescale can be studied in terms of the burst size distribution~\cite{Karsai2012Universal}. More recently, a burst tree representation method has been suggested to fully characterize temporal correlations in the time series for the entire range of timescale~\cite{Jo2020Bursttree}. In addition, the conventional, autocorrelation function (ACF) has also been employed to detect long-range temporal correlations in the bursty time series.

However, the above mentioned methods tend to have focused on a single time series derived from a temporal network. For example, an event sequence can be derived from a set of events occurred on multiple links adjacent to a node of interest. In general, event sequences can be defined on any subset of links in the temporal network. By aggregating the events over a subset of links, information on the topological structure within the subset is missing. This issue strongly calls for a systematic approach to the detection of temporal correlations in temporal networks. There have been several approaches based on the dynamical processes on temporal networks~\cite{Karsai2011Small, Rocha2011Simulated, Kivela2012Multiscale, Delvenne2015Diffusion}. Recently, methods of detecting recurrent states of temporal networks have been suggested based on similarity or dissimilarity measures between two snapshot networks separated by some time interval~\cite{Masuda2019Detecting, Cao2020Detecting, Sugishita2021Recurrence}. One of such dissimilarity measures is the network distance, and it is used to define the ACF for temporal networks~\cite{Sugishita2021Recurrence}.

In this work we focus on analysis of the ACF for temporal networks by assuming that activity patterns on links are independent of each other and that links have heterogeneous levels of activity. For this, we first define the ACF for temporal networks in discrete time, which is similar to but different from the definition in Ref.~\cite{Sugishita2021Recurrence}. Then we derive an approximately analytical form of the ACF by employing the copula method developed for a single time series~\cite{Jo2019Analytically}. The analytical solution is also compared to the numerical simulations, for which a copula-based algorithm for generating bursty time series is used after some modification~\cite{Jo2019Copulabased}.

\section{Analysis}

\subsection{Definition of the autocorrelation function for temporal networks}

We define an autocorrelation function (ACF) for temporal networks. For this, we consider a temporal network with $N$ nodes and $L$ links defined in discrete times of $t=0,\ldots,T-1$. The activity pattern on each link $l=1,\ldots,L$ can be described by a sequence of interaction events occurred on the link $l$, or an event sequence. Equivalently, it can be described by a time series $x_l(t)$ that has a value of $1$ when the event occurs at the time step $t$, $0$ otherwise. The state of the network at the time step $t$, or a snapshot network, is denoted by a vector $\vec x(t)=(x_1(t),\ldots, x_L(t))$. Then, the ACF of the network with time delay $t_{\rm d}$ is defined as
\begin{align}
    A(t_{\rm d})\equiv \frac{\langle \vec x(t)\cdot \vec x(t+t_{\rm d})\rangle_t - \langle \vec x(t)\rangle_t\cdot \langle \vec x(t)\rangle_t}{\langle \vec x(t)\cdot \vec x(t)\rangle_t - \langle \vec x(t)\rangle_t\cdot \langle \vec x(t)\rangle_t},
    \label{eq:acf_define}
\end{align}
where $\langle\cdot\rangle_t$ denotes the time average of a variable over the period of $0\leq t\leq T-1$ whether the variable is a scalar or a vector. For any $t$ and $t_{\rm d}$, the inner product $\vec x(t)\cdot \vec x(t+t_{\rm d})=\sum_{l=1}^L x_l(t)x_l(t+t_{\rm d})$ is the number of links in which events occur both at time steps $t$ and $t+t_{\rm d}$. Note that the definition of the ACF in Ref.~\cite{Sugishita2021Recurrence} misses the second term in the numerator of the right hand side in Eq.~\eqref{eq:acf_define}. By our definition, $A(0)=1$, hence we consider $t_{\rm d}>0$ unless otherwise stated hereafter.

\subsection{Copula-based analysis}

For a given event sequence on a link $l$, one can get the IET distribution $P_l(\tau)$. If $n_l$ events occur on a link $l$ for the period of $[0,T-1]$, the event rate is calculated as 
\begin{align}
    \lambda_l\equiv \langle x_l(t)\rangle_t=n_l/T\simeq 1/\mu_l,
\end{align}
where $\mu_l$ is an average of IETs and $\lambda_l=1/\mu_l$ in the limit of $n_l,T\to \infty$. Thus, the term $\langle \vec x(t)\rangle_t$ in Eq.~\eqref{eq:acf_define} reads
\begin{align}
    \langle \vec x(t)\rangle_t=\left(\lambda_1,\ldots, \lambda_L\right),
\end{align}
leading to
\begin{align}
    \langle \vec x(t)\rangle_t \cdot \langle \vec x(t)\rangle_t =\sum_{l=1}^L \lambda_l^2.
\end{align}
One also obtains
\begin{align}
    \langle \vec x(t)\cdot \vec x(t)\rangle_t =  \sum_{l=1}^L \langle x^2_l(t)\rangle_t = \sum_{l=1}^L \langle x_l(t)\rangle_t = \sum_{l=1}^L \lambda_l,
\end{align}
where we have used the fact that $x^2_l(t)=x_l(t)$ as $x_l(t)=0,1$. We get
\begin{align}
    \langle \vec x(t)\cdot \vec x(t+t_{\rm d})\rangle_t = \sum_{l=1}^L \langle x_l(t)x_l(t+t_{\rm d})\rangle_t,
\end{align}
where each term in the summation can be written as
\begin{align}
    \langle x_l(t)x_l(t+t_{\rm d})\rangle_t = \lambda_l \sum_{k=1}^\infty P_{lk}(t_{\rm d}).
\end{align}
Here $P_{lk}(t_{\rm d})$ is the probability that two events occurred at time steps $t$ and $t+t_{\rm d}$ on the link $l$ are separated by exactly $k$ IETs for a positive integer $k$~\cite{Jo2019Analytically}. We assume that the properties of $P_{lk}(t_{\rm d})$ are independent of those of other links by ignoring possible link-link correlations~\cite{Saramaki2015Seconds, Hiraoka2020Modeling, Williams2022NonMarkovian}. Assuming a continuous time for the analytical tractability, $P_{lk}(t_{\rm d})$ can be written as
\begin{align}
    P_{lk}(t_{\rm d})=\prod_{i=1}^k \int_0^\infty d\tau_i P_l(\tau_1,\ldots,\tau_k)\delta\left(t_{\rm d}-\sum_{i=1}^k \tau_i\right),
    \label{eq:Plktd}
\end{align}
where $P_l(\tau_1,\ldots,\tau_k)$ is a joint probability distribution of $k$ consecutive IETs for the link $l$ and $\delta(\cdot)$ is a Dirac delta function. Then the ACF $A(t_{\rm d})$ in Eq.~\eqref{eq:acf_define} is written as
\begin{align}
    A(t_{\rm d})=\frac{\sum_{l=1}^L \lambda_l \sum_{k=1}^\infty P_{lk}(t_{\rm d}) -\sum_{l=1}^L \lambda_l^2}{\sum_{l=1}^L \lambda_l -\sum_{l=1}^L \lambda_l^2}.
    \label{eq:acf_simple}
\end{align}

We remark that the joint probability distribution $P_l(\tau_1,\ldots,\tau_k)$ carries information on the correlation structure between consecutive IETs for a link $l$. In our work we consider the simplest correlation structure such that each IET is conditioned only by its previous IET, which can be interpreted as a Markovian property. The correlation between two consecutive IETs is then characterized by the memory coefficient $M_l$ for a link $l$~\cite{Goh2008Burstiness}. The memory coefficient $M_l$ is defined as a Pearson correlation coefficient between two consecutive IETs, say $\tau_i$ and $\tau_{i+1}$, as follows:
\begin{align}
    M_l\equiv \frac{\langle \tau_i\tau_{i+1}\rangle-\mu_l^2}{\sigma_l^2},
    \label{eq:M_define}
\end{align}
where 
\begin{align}
    \langle \tau_i\tau_{i+1}\rangle\equiv \int_0^\infty d\tau_i\int_0^\infty d\tau_{i+1} \tau_i \tau_{i+1} P_l(\tau_i, \tau_{i+1}),
\end{align}
and $\mu_l$ and $\sigma_l$ are the mean and standard deviation of $P_l(\tau)$, respectively.

Thanks to the assumed Markovian property, one can factorize the joint probability distribution $P_l(\tau_1,\ldots,\tau_k)$ as follows:
\begin{align}
    P_l(\tau_1,\ldots,\tau_k) = \prod_{i=1}^{k-1} P_l(\tau_i, \tau_{i+1}) \Big/ \prod_{i=2}^{k-1} P_l(\tau_i).
    \label{eq:jointPDF}
\end{align}
For modeling the bivariate probability distribution $P_l(\tau_i, \tau_{i+1})$ we employ a Farlie-Gumbel-Morgenstern (FGM) copula among many others~\cite{Nelsen2006Introduction, Takeuchi2010Constructing} by assuming that
\begin{align}
    P_l(\tau_i,\tau_{i+1}) = P_l(\tau_i) P_l(\tau_{i+1}) \left[1+r_l f_l(\tau_i)f_l(\tau_{i+1})\right],
    \label{eq:FGM_define}
\end{align}
where
\begin{align}
    f_l(\tau) \equiv 2F_l(\tau)-1,\ F_l(\tau) \equiv \int_0^{\tau} d\tau' P_l(\tau').
\end{align}
Here the parameter $r_l\in[-1,1]$ controls the degree of correlation between two consecutive IETs. It is straightforward to prove that the parameter $r_l$ is proportional to the memory coefficient $M_l$ in Eq.~\eqref{eq:M_define}:
\begin{align}
    M_l =\frac{r_l}{\sigma_l^2} \left[ \int_0^\infty d\tau \tau P_l(\tau)f_l(\tau) \right]^2 \equiv a_lr_l.
    \label{eq:Mr_relation}
\end{align}
Plugging Eq.~\eqref{eq:FGM_define} into Eq.~\eqref{eq:jointPDF} one gets
\begin{align}
    P_l(\tau_1,\cdots,\tau_k) = \prod_{i=1}^k P_l(\tau_i) \prod_{i=1}^{k-1} \left[1+r_l f_l(\tau_i) f_l(\tau_{i+1})\right]. 
\end{align}
By assuming that $|r_l|\ll 1$, we expand the above equation up to the first order of $r_l$ as follows:
\begin{align}
    P_l(\tau_1,\cdots,\tau_k) \approx \prod_{i=1}^k P_l(\tau_i) \left[1+ r_l \sum_{i=1}^{k-1} f_l(\tau_i)f_l(\tau_{i+1}) + \mathcal{O}(r_l^2)\right].
    \label{eq:Pltaus_approx}
\end{align}
Using Eq.~\eqref{eq:Pltaus_approx} we take the Laplace transform of $P_{lk}(t_{\rm d})$ in Eq.~\eqref{eq:Plktd} to get
\begin{align}
    &\widetilde{P_{lk}}(s) \equiv \int_0^\infty dt_{\rm d} P_{lk}(t_{\rm d}) e^{-st_{\rm d}}\nonumber\\
    &\approx \tilde{P_l}(s)^k + r_l(k-1)\tilde{P_l}(s)^{k-2}\tilde{Q_l}(s)^2 + \mathcal{O}(r_l^2),
\end{align}
where
\begin{align}
    &\tilde{P_l}(s) \equiv \int_0^\infty d\tau P_l(\tau)e^{-s\tau},\\
	&\tilde{Q_l}(s) \equiv \int_0^\infty d\tau P_l(\tau) f_l(\tau) e^{-s\tau}.
\end{align}
One obtains up to the first order of $r_l$
\begin{align}
    \sum_{k=1}^\infty \widetilde{P_{lk}}(s) \approx \frac{ \tilde{P_l}(s) }{ 1 - \tilde{P_l}(s) } + \frac{r_l \tilde{Q_l}(s)^2}{[1-\tilde{P_l}(s)]^2} + \mathcal{O}(r_l^2).
	\label{eq:Plk_s_approx}
\end{align}
To get the ACF in Eq.~\eqref{eq:acf_simple}, one needs to take the inverse Laplace transform of Eq.~\eqref{eq:Plk_s_approx} for given $P_l(\tau)$ for each $l$, and then to plug them into Eq.~\eqref{eq:acf_simple}.

\subsection{Case with exponential IET distributions}

We study the case with an exponential IET distribution for all links but with different levels of activity $\lambda$:
\begin{align}
    P_l(\tau)=\lambda_le^{-\lambda_l\tau}.
    \label{eq:Pltau}
\end{align}
Note that for the exponential IET distribution, $a_l=1/4$ in Eq.~\eqref{eq:Mr_relation} regardless of $\lambda_l$, implying that $r_l=4M_l$ for all $l$s. Here we assume that $r_l=r$, hence $M_l=M$, for all $l$s, and that $\lambda_l$ is uniformly distributed over $(0,1]$, i.e., 
\begin{align}
    P(\lambda)=1\ \textrm{for}\ \lambda\in(0,1], 
    \label{eq:Plambda}
\end{align}
implying the power-law distribution of $\mu$ values:
\begin{align}
    P(\mu)=\mu^{-2}\ \textrm{for}\ \mu\geq 1.
\end{align}
Then for each link $l$, we get
\begin{align}
    \sum_{k=1}^\infty P_{lk}(t_{\rm d})\approx \lambda_l +4M\lambda_l^2 t_{\rm d} e^{-2\lambda_l t_{\rm d}}+ \mathcal{O}(M^2),
\end{align}
enabling us to get from Eq.~\eqref{eq:acf_simple}
\begin{align}
    A(t_{\rm d})\approx \frac{4Mt_{\rm d} \langle \lambda_l^3 e^{-2\lambda_l t_{\rm d}} \rangle}{\langle \lambda_l \rangle - \langle \lambda_l^2 \rangle} +\mathcal{O}(M^2),
\end{align}
where $\langle \cdot\rangle$ is the ensemble average using $P(\lambda)$ in Eq.~\eqref{eq:Plambda}. Since
\begin{align}
    \langle \lambda_l \rangle=\frac{1}{2},\ \langle \lambda_l^2 \rangle=\frac{1}{3},
\end{align}
one finally obtains
\begin{align}
    A(t_{\rm d})\approx \frac{3M\left[3-(4t_{\rm d}^3+6t_{\rm d}^2++6t_{\rm d}+3)e^{-2t_{\rm d}}\right]}{t_{\rm d}^3}+\mathcal{O}(M^2).
    \label{eq:acf_result}
\end{align}
Note that for $t_{\rm d}\gg 1$, $A(t_{\rm d})\approx 9Mt_{\rm d}^{-3}$, implying that the power-law decaying behavior of the ACF may appear in temporal networks with heterogeneous link activities even when the IETs are exponentially distributed in each link of the network.

\section{Numerical simulation}

To numerically validate our analytical solution in Eq.~\eqref{eq:acf_result} we perform numerical simulations by generating $L$ event sequences with different values of the event rate $\lambda$. To realize $P(\lambda)=1$ for $\lambda\in(0,1]$ in Eq.~\eqref{eq:Plambda} we set the value of $\lambda_l$ as follows:
\begin{align}
    \lambda_l=\frac{(L-2)l+1}{L(L-1)}\ \textrm{for}\ l=1,\ldots,L.
    \label{eq:lambda_l}
\end{align}
That is, $\lambda_1=1/L$ and $\lambda_L=1-1/L$, which is to avoid two extreme cases of $\lambda=0$ and $1$. The case with $\lambda=0$ means no events at all, while $\lambda=1$ indicates the case when events occur in every time step.

\begin{figure}[!b]
\includegraphics[width=\columnwidth]{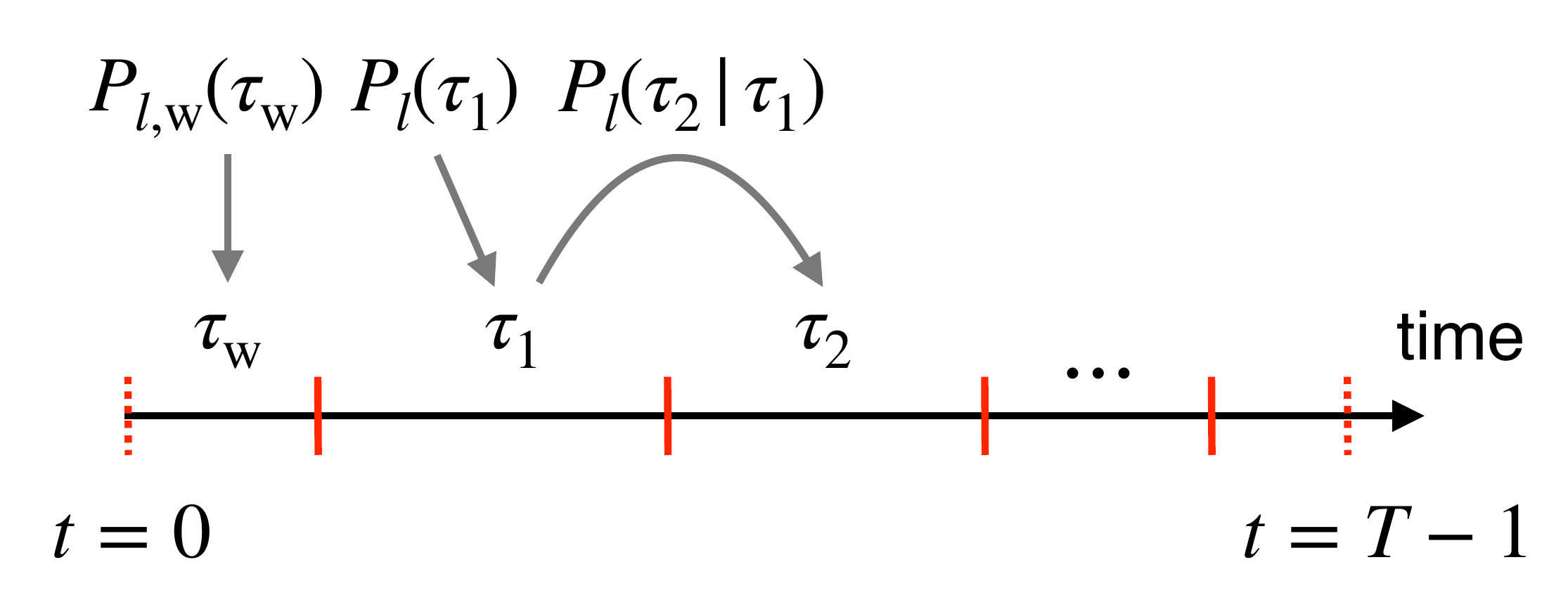}
\caption{Schematic diagram for generating an event sequence with correlated interevent times on the $l$th link of the temporal network. See the main text for definitions of symbols.}
\label{fig:simulation}
\end{figure}

\begin{figure}[!t]
\includegraphics[width=0.8\columnwidth]{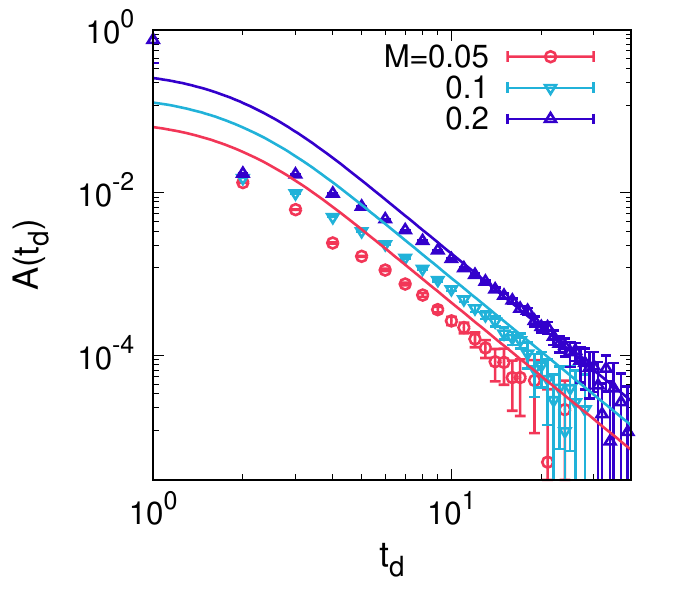}
\caption{Simulation results (symbols) of the autocorrelation function $A(t_{\rm d})$ in Eq.~\eqref{eq:acf_numerical} for several values of the memory coefficient $M$, compared with the corresponding analytical solution in Eq.~\eqref{eq:acf_result} (solid lines). Each numerical curve is averaged over $50$ temporal networks generated using $L=10^3$ and $T=10^5$ for uniformly distributed $\lambda_l$s in Eq.~\eqref{eq:lambda_l}. Error bars denote the standard errors.}
\label{fig:acf_numerical}
\end{figure}

To generate bursty time series with correlated interevent times (IETs) we employ the copula-based algorithm introduced in Ref.~\cite{Jo2019Copulabased} but with a modification: in order to make the temporal network ``stationary'' from $t=0$, we introduce a residual time in the beginning~\cite{Jo2014Analytically}, see the schematic diagram in Fig.~\ref{fig:simulation}. The residual time distribution for a link $l$, denoted by $P_{l,\rm w}(\tau_{\rm w})$, is directly obtained from the IET distribution for a link $l$ as follows:
\begin{align}
    P_{l,\rm w}(\tau_{\rm w})=\frac{1}{\mu_l}\int_{\tau_{\rm w}}^\infty P_l(\tau)d\tau,
\end{align}
from which we draw a value $\tau_{\rm w}$ to set the timing of the first event on a link $l$ as $t_1=\tau_{\rm w}$. Then, the second event occurs in time $t_2=t_1+\tau_1$, where the IET $\tau_1$ is drawn from $P_l(\tau)$. Given the $i$th IET $\tau_i$, the next IET $\tau_{i+1}$ is drawn from the conditional probability distribution:
\begin{align}
    P_l(\tau_{i+1}|\tau_i) = P_l(\tau_{i+1}) \left[1+r_l f_l(\tau_i)f_l(\tau_{i+1})\right].
    \label{eq:condPDF}
\end{align}
Using the sequence of generated IETs $\left\{\tau_i\right\}$, one obtains the sequence of event timings, i.e., $t_j=\tau_{\rm w}+\sum_{i=1}^{j-1} \tau_i$ for $j=1,\ldots,\tilde n_l$. Here $\tilde n_l$ is the number of events in the event sequence, which is determined by the condition that $t_{\tilde n_l}\leq T-1$ and $t_{\tilde n_l+1}\geq T$. We remark that since the minimum IET is $0$ by definition of $P_l(\tau)$ in Eq.~\eqref{eq:Pltau}, we add one to each drawn IET and then round event timings to make them discrete, which may cause deviations of the numerical simulations from the analytical solution. By collecting the $L$ event sequences generated for $L$ links, one finally gets a temporal network.

Once the temporal network is generated, the autocorrelation function (ACF) is calculated using a numerical version of the ACF defined in Eq.~\eqref{eq:acf_define}:
\begin{align}
    A(t_{\rm d})=\frac{\sum_{l=1}^L\left[ \frac{1}{T-t_{\rm d}} \sum_{t=0}^{T-t_{\rm d}-1} x_l(t)x_l(t+t_{\rm d}) -\lambda_{l,1}\lambda_{l,2} \right]}{\sum_{l=1}^L \sigma_{l,1}\sigma_{l,2}},
    \label{eq:acf_numerical}
\end{align}
where $\lambda_{l,1}$ and $\sigma_{l,1}$ are respectively the average and standard deviation of $x_l(t)$ for $t=0,\ldots, T-t_{\rm d}-1$, while $\lambda_{l,2}$ and $\sigma_{l,2}$ are respectively the average and standard deviation of $x_l(t)$ for $t=t_{\rm d},\ldots, T-1$. 

We generate $50$ temporal networks using $L=10^3$ and $T=10^5$ for each value of $M=0.05$, $0.1$, and $0.2$. Then we obtain numerical results of the ACF in Eq.~\eqref{eq:acf_numerical}, denoted by symbols in Fig.~\ref{fig:acf_numerical}. We find that these numerical results are comparable with the analytical solution in Eq.~\eqref{eq:acf_result} that are denoted by solid lines in Fig.~\ref{fig:acf_numerical}. The systematic deviation of the numerical results from the analytical solution might be due to the higher-order terms of $r$ or $M$ ignored in the analytical derivation as well as the numerical errors in the simulation. In particular, the deviations for small $t_{\rm d}$ might be induced by rounding errors in generating discrete event timings from continuous IET distributions, as mentioned above.

\section{Conclusion}

To characterize temporal correlations in temporal networks, we first define an autocorrelation function (ACF) for temporal networks in terms of the similarity between two snapshot networks separated by a certain time interval. Then by employing a copula-based method developed for a single time series~\cite{Jo2019Analytically} we derive the ACF for a temporal network in which activity patterns of links are independent of each other but their activity levels are heterogeneous. By assuming that exponential distributed interevent times (IETs) are weakly correlated with each other, we finally derive an analytical solution of the ACF. The validity of the analytical solution is tested against the numerical simulations, for which we adopt the copula-based algorithm for generating bursty time series~\cite{Jo2019Copulabased}, to find that the numerical results are comparable to the analytical solution.

This work can be extended to consider alternative definitions of the ACF based on other network distance indexes summarized in Ref.~\cite{Masuda2019Detecting}, e.g., a graph edit distance~\cite{Gao2010Survey}. One can incorporate more realistic, heavy-tailed IET distributions, despite the fact that it is highly non-trivial to analyze the ACF as partly shown in Ref.~\cite{Jo2019Analytically}. Moreover, correlations between more than two consecutive IETs can be considered by means of vine copula methods~\cite{Takeuchi2020Constructing, Jo2022Copulabased}. Since the link-link correlations are important to understand the empirical temporal networks~\cite{Saramaki2015Seconds, Williams2022NonMarkovian}, one can also take link-link correlations into account for the analysis of the ACF. Finally, the ACF defined in our work can be calculated for empirical temporal networks to see if such long-term temporal correlations are present in reality.

\begin{acknowledgments}
H.-H.J. acknowledges financial support by the National Research Foundation of Korea (NRF) grant funded by the Korea government (MSIT) (No. 2022R1A2C1007358). 
\end{acknowledgments}



%

\end{document}